\documentclass{article} % For LaTeX2e
\usepackage{iclr2024_conference,times}

\usepackage{amsmath,amsfonts,bm}

\def\eqref#1{equation~\ref{#1}}
\def\1{\bm{1}}

\DeclareMathAlphabet{\mathsfit}{\encodingdefault}{\sfdefault}{m}{sl}
\SetMathAlphabet{\mathsfit}{bold}{\encodingdefault}{\sfdefault}{bx}{n}

\usepackage{hyperref}
\usepackage{url}
\usepackage{amsmath}
\usepackage{amssymb}
\usepackage{mathtools}
\usepackage{amsthm}
\usepackage{graphicx} % For including images
\usepackage{caption} % For captions
\usepackage{subcaption} % For subfigures
\usepackage{algorithm}
\usepackage{algpseudocode}
\usepackage{multirow}
\usepackage{microtype}
\usepackage{booktabs} % for professional tables
\usepackage{float}
\usepackage{ifthen}

\title{Accelerating the Generation of Molecular Conformations with Progressive Distillation of Equivariant Latent Diffusion Models}

\author{Romain Lacombe \\
  Stanford University\\
  \texttt{rlacombe@stanford.edu} \\
    \And
  Neal Vaidya \\
  NVIDIA\\
  \texttt{nealv@nvidia.com} \\
}

\iclrfinalcopy % Uncomment for camera-ready version, but NOT for submission.
\begin{document}

\maketitle

\begin{abstract}
  Recent advances in fast sampling methods for diffusion models have demonstrated significant potential to accelerate generation on image modalities. We apply these methods to 3-dimensional molecular conformations by building on the recently introduced GeoLDM  equivariant latent diffusion model \citep{xu2023geometric}. We evaluate trade-offs between speed gains and quality loss, as measured by molecular conformation structural stability. We introduce Equivariant Latent Progressive Distillation, a fast sampling algorithm that preserves geometric equivariance and accelerates generation from latent diffusion models. Our experiments demonstrate \textbf{up to 7.5$\times$ gains in sampling speed} with limited degradation in molecular stability. These results suggest this accelerated sampling method has strong potential for high-throughput \emph{in silico} molecular conformations screening in computational biochemistry, drug discovery, and life sciences applications.
\end{abstract}

\section{Introduction}

Generative approaches in scientific domains have the potential to model complex physical systems and processes without having to solve often intractable physical equations from first principles. Recent work on geometric latent diffusion-based generative models (GeoLDM, \citet{xu2023geometric}) has demonstrated state-of-the-art results in structural accuracy of 3-dimensional molecular conformations. However, diffusion models require iterative inference, which slows down generation significantly. This limits the tractability of diffusion-based generative models for high-throughput \emph{in silico} screening of large bio-molecules, which are of critical importance to biochemistry and life sciences.

Recent advances on faster sampling of diffusion models \citep{salimans2022progressive, song2023consistency, luo2023latent, shih2023parallel, zheng2023fast, meng2023distillation} have demonstrated significant generation speed gains for image modalities and robotics tasks, while maintaining high degrees of quality. In this project, we apply recent methods for accelerated sampling of diffusion models to the 3-dimensional molecular conformations generation task, which is of high importance for ligand docking and virtual screening in computational drug discovery. Our objective is to accelerate the generation of physically plausible molecular conformations by diffusion models. 

\textbf{Our core research question: what are the speed vs. quality trade-offs for accelerated sampling of molecular conformations with geometric latent diffusion models (GeoLDM), and which methods provide the most speed gain for the least quality loss?}

Given the demonstrated success of diffusion acceleration techniques in speeding up diffusion models for image modalities, both in wall time and number of inference steps, we expect their application to molecular conformations generation to lead to significant speed gains, albeit at the expense of some loss of quality. We implement accelerated diffusion methods for molecular conformation generation, and evaluate these techniques against the GeoLDM baseline. We specifically report on speed/quality trade-offs for each acceleration strategy, as measured by sampling speed improvements vs. structural stability degradation for generated conformations.

%%%%%%%%%%%%%%%%%%%%%%%%%%%%%%%%%%%%%%%%%%%%%%%%

\section{Related Work}

Several recent papers have demonstrated significant potential to accelerate generation from diffusion models by parallelizing inference or learning to sample multiple (or all) diffusion steps at once.

\textbf{Parallel sampling of diffusion models.} \citet{shih2023parallel} demonstrate in recent work a 2-4x improvement in sampling speed across a range of robotics and image generation tasks, by training a diffusion model to take several steps at once in parallel. Importantly, these speed gains did not result in measurable loss of quality assessed against various quantitative generation evaluation benchmarks.

\textbf{Fast sampling of diffusion models via operator learning.} \citet{zheng2023fast} apply operator learning to the fast sampling problem, and solve the induced (deterministic) probability flow ordinary differential equation (ODE) from score-based generative modeling through stochastic differential equations (SDE) using a neural operator. By mapping initial conditions (random sample) to the full trajectory that solves the probability flow ODE, where the last point in the trajectory is the data, this approach parallelizes sampling at each time step and allows for fast sampling from a diffusion process in a single model forward step.

\textbf{Consistency models.} To address the slow, iterative generation process of diffusion-based models, \cite{song2023consistency} and \cite{luo2023latent} introduce consistency models, which are trained to directly map noise to data in a single step, either by distilling pre-trained diffusion models or through direct, end-to-end training. Their recent results surpass existing diffusion model distillation techniques and achieve state-of-the-art performance on image modality benchmarks.

%%%%%%%%%%%%%%%%%%%%%%%%%%%%%%%%%%%%%%%%%%%%%%%%

\section{Methods}

%***************************
\begin{figure}[!b]
  \centering
      \includegraphics[width=\linewidth]{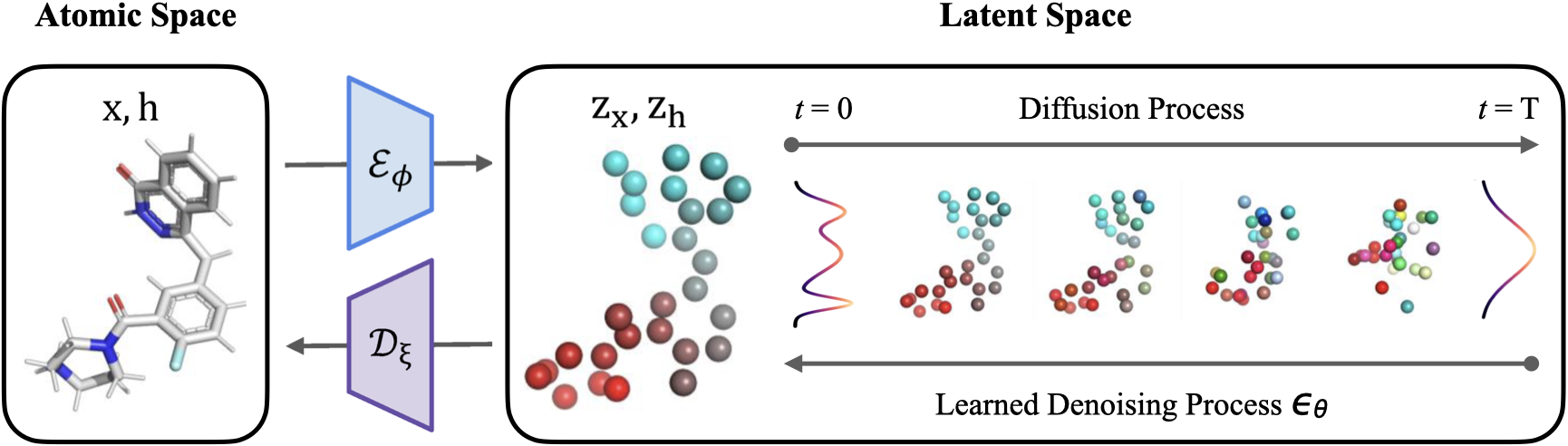}
  \caption{\textbf{Geometric latent diffusion model.} In the GeoLDM model \citep{xu2023geometric}, an equivariant denoising process is trained to generate point clouds in latent space by inverting a diffusion process. A decoder then transforms point clouds into 3-dimensional molecular conformations in atomic space. Image adapted from \citet{xu2023geometric} and \citet{neurips23diffusionworkshop}.}
  \label{fig:GeoLDM}
\end{figure}
%***************************
 
\subsection{Geometric Latent Diffusion Models (GeoLDM)} 
Our paper builds on recent work in molecular conformation generation with equivariant latent diffusion models \citep{xu2023geometric, hoogeboom2022equivariant}. In the GeoLDM framework (Figure \ref{fig:GeoLDM}), an encoder learns a latent representation of atom-space molecular conformations as a point cloud in latent space. The denoising model learns to invert a diffusion process in latent space, to generate point cloud representations of plausible conformations through successive inference steps, starting from Gaussian noise. A decoder then transforms the output of this latent diffusion denoising process into molecular conformations in physical space.

Importantly, the original Progressive Distillation paper \citep{salimans2022progressive} accelerates diffusion models in physical space. Here, we introduce and experiment with progressive distillation of the diffusion denoising process of equivariant latent diffusion models, directly in latent space. We experiment with sampling from the teacher model both deterministically (DDIM) and stochastically (DDPM), as explained below.

%%%%%%%%%%%%%%%%%%%%%%%%

\subsection{Denoising Diffusion Implicit Model (DDIM)}

Denoising Diffusion Implicit Models (DDIM, \citet{song2022denoising}) offer a computationally efficient alternative to traditional diffusion models through an implicit denoising process that allows for larger jumps in the state space, significantly reducing the number of required iterative steps.

In a standard Denoising Diffusion Probabilistic Model (DDPM, \citet{ho2020denoising}), the data sampling procedure is defined as a stochastic Markov process in which some $z_T$ sampled from the prior distribution $p(\mathbf{z}_T)$ is iteratively denoised by sampling from $p(\mathbf{z}_{t-1}|\mathbf{z}_{t}, t)$ to produce the output $\mathbf{z}_0$. This poses challenges for acceleration and approximation, since each $\mathbf{z}_{t-1}$ relies only on $\mathbf{z}_t$ and cannot be approximated by using previous values. 

We can instead define a different, non-Markovian generative sampling process: 
$$
    \mathbf{z}_{t-1} = \sqrt{\alpha_{t-1}} \mathbf{\hat{z}}_0 +
        \sqrt{1 - \alpha_{t-1} - \sigma_t^2} \cdot \epsilon_\theta^{(t)}(\mathbf{z}_t) + 
        \sigma_t \epsilon_t
$$
where $\epsilon_t \sim \mathcal{N}(0,I)$, $\epsilon_\theta^{(t)}(\mathbf{z}_t)$ is a prediction of the noise from the first $t$ steps of the forward noising process, and $\mathbf{\hat{z}}_0$ is an approximation of $\mathbf{z}_0$. Since this process no longer relies only on the value of $\textbf{z}_t$, we can approximate $\mathbf{z}_{t-1}$ without first explicitly sampling $\mathbf{z}_t$. \cite{song2022denoising} showed that this alternative sampling process is equivalent to the DDPM process when: 
$$ \mathbf{\hat{z}}_0 = \frac{\mathbf{z}_t - \sqrt{1-\alpha_t}\epsilon_\theta^{(t)}(\mathbf{z}_t)}{\sqrt{\alpha_t}}
\quad \quad \sigma_t = \sqrt{\frac{(1-\alpha_{t-1})}{1-\alpha_t}}\sqrt{\frac{1-\alpha_t}{\alpha_{t-1}}} $$

Setting $\sigma_t = 0$ for all $t$ leads to a fully deterministic procedure to sample from a Denoising Diffusion \emph{Implicit} Model (DDIM). Non-stochastic denoising opens up new avenues for accelerating sampling by approximating the deterministic trajectory, either through skipping certain sample steps $\mathbf{z}_t$ or by training a new model to match the denoising trajectory of the original model. 

%%%%%%%%%%%%%%%%%%%%%%%%

\subsection{Equivariant Latent Progressive Distillation}

%***************************
\begin{figure}[b]
  \centering
      \includegraphics[width=\linewidth]{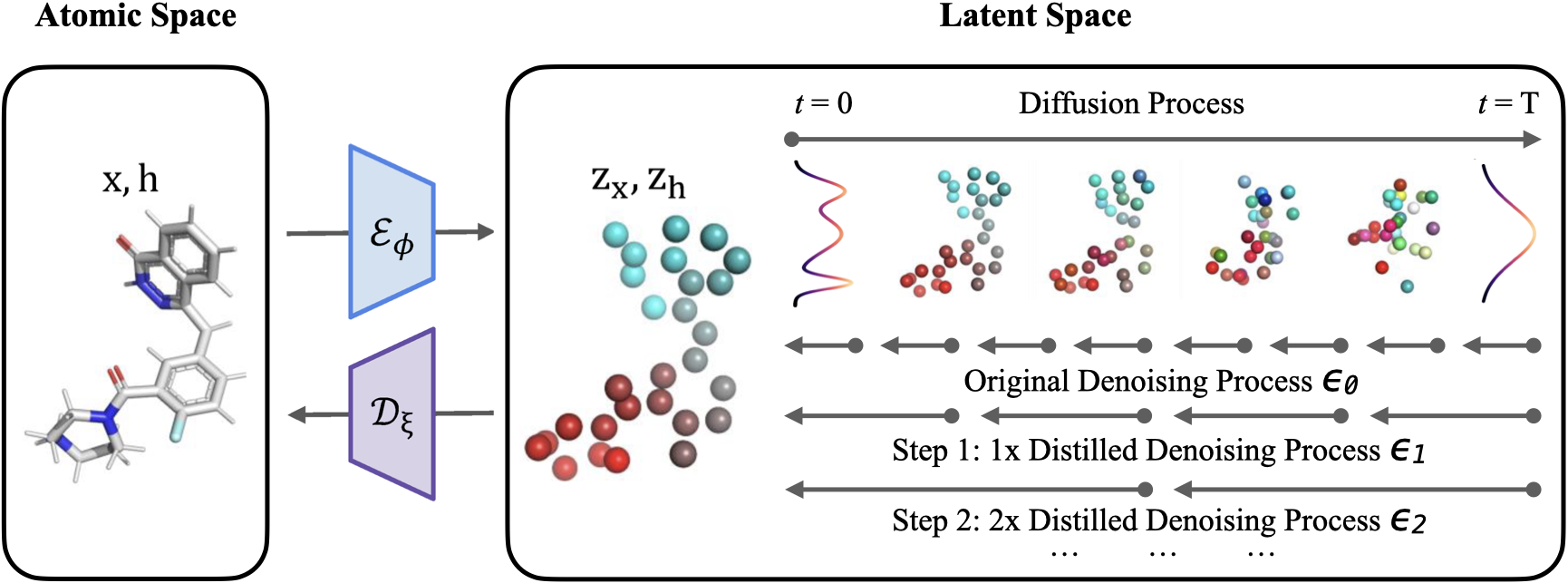}
  \caption{\textbf{Equivariant latent progressive distillation:} the denoising process in latent space is iteratively distilled by teaching a student model to sample two steps at a time from the teacher model. The algorithm initializes from the original denoising model, and progresses by successive halvings.}
  \label{fig:progressive-distillation}
\end{figure}
%***************************

In the progressive distillation approach, a trained diffusion model is distilled through an iterative student-teacher framework where the student model is trained to output in one denoising step a sample corresponding to two denoising steps of the teacher model. This distilled model can generate samples comparable to those generated by the teacher model, but with only half the required diffusion steps, a 2$\times$ speed-up. The process can then be repeated with successive distillation steps.

\textbf{Here we introduce the Equivariant Latent Progressive Distillation (ELPD) algorithm}, which applies progressive distillation to the diffusion denoising process of an equivariant latent diffusion model directly in latent space. The Equivariant Latent Progressive Distillation algorithm is presented in Figure \ref{fig:progressive-distillation}, and in algorithm \ref{alg:progdist} in appendix.

We implement Equivariant Latent Progressive Distillation in the GeoLDM framework, starting from the original GeoLDM model weights pre-trained on \emph{QM9} released by \citet{xu2023geometric}. Following the implementation in \cite{salimans2022progressive}, we set the weighting function $w(\lambda_t)$ to be the Truncated SNR, defined as $ \max(\alpha_t^2 / \sigma_t^2, 1)$. We experiment with both deterministic sampling of latent diffusion models using Denoising Diffusion Implicit Models (DDIM), and with progressive distillation of equivariant diffusion models in latent space using both deterministic DDIM sampling (DDIM-PD) and stochastic Denoising Diffusion Probabilistic Models sampling (DDPM-PD).

To our knowledge, this is the first implementation of a progressive distillation setup in latent space for equivariant diffusion generative models of molecular conformations.

%%%%%%%%%%%%%%%%%%%%%%%%%%%%%%%%%%%%%%%%

\section{Evaluation}

%%%%%%%%%%%%%%%%%%%%%%%%
\subsection{Dataset}

Consistent with our baseline model GeoLDM \citep{xu2023geometric}, we train on the \emph{QM9} small molecules dataset \citep{ramakrishnan2014QM9}, a comprehensive dataset of 134,000 stable and synthetically accessible small organic molecules, each containing up to 9 non-hydrogen atoms of carbon (C), oxygen (O), nitrogen (N), and fluorine (F) (up to 29 atoms when accounting for hydrogen H). \emph{QM9} provides stable 3-dimensional molecular conformation structures, as well as a set of quantum chemistry features including geometric, energetic, electronic, and thermodynamic properties.

%%%%%%%%%%%%%%
\subsection{Performance Metrics: Speed vs. Quality}

\paragraph{Speed:}Sampling speed is reported both as the number of iterations at sample time (diffusion steps) and as the number of samples generated per second, averaged over the generation of 10,000 samples. Except for models with very small numbers of diffusion steps, where model overhead makes each step relatively slower, we find that wall-time generation speed and number of diffusion steps are in almost perfect correlation with one another.

\paragraph{Quality:}Following GeoLDM \citep{xu2023geometric}, we evaluate the quality of our models by assessing chemical plausibility of the molecular conformations generated by sampling on the \emph{QM9} dataset. We report the following performance metric for each evaluation run: 
\begin{itemize}
    \item \textbf{Atom-level stability:} proportion of atoms that are stable, i.e. generated with correct valence. 
    \item \textbf{Molecule-level stability:} proportion of generated molecules where all atoms are stable. 
    \item \textbf{Validity:} proportion of molecules measured as valid using \textsc{RDKit} \citep{RDKit}. 
    \item \textbf{Uniqueness:} the percentage of unique molecules in all generated conformations. 
      
\end{itemize}

%%%%%%%%%%%%%%%%%%%%%%%%%%%%%%%%%%%%%%%%%%%%%%%%

\subsection{Experiments}

We report sample quality and generation speed for the following baseline models:
\begin{itemize}
\item \textbf{GeoLDM-1000} equivariant latent diffusion model (in latent space)  \cite{xu2023geometric} pre-trained with the original $T=1000$ diffusion steps;
\item \textbf{GeoLDM-100} trained with a reduced number of diffusion steps $T=100$. 
\end{itemize}

We then compare performance metrics for these baselines with the following experiments: 

\begin{itemize}

\item \textbf{Denoising diffusion implicit model with reduced diffusion steps:} we sample from GeoLDM as a DDIM model with $T=\{1000, 500, 250, 125, 63, 32, 16, 8, 4, 2, 1\}$ steps.

\item \textbf{Latent equivariant progressive distillation with deterministic sampling:} we train successive student models through progressive distillation, with $T=\{1000, 500, 250, 125, 63, 32, 16\}$ diffusion steps, and use deterministic sampling (DDIM) at train time.

\item \textbf{Latent equivariant progressive distillation with stochastic sampling:} we train successive student models through progressive distillation, with $T=\{1000, 500, 250, 125, 63, 32, 16\}$ diffusion steps, and use stochastic sampling (DDPM) at train time.
\end{itemize}

We report results for baselines and all experiments in Table \ref{tab:comparison} and Figure \ref{fig:comparison}. We also report examples of visualizations of conformations generated by our models in appendix (Figure \ref{fig:visualizations}).

\captionsetup[table]{skip=10pt}
\begin{table*}[ht]
\begin{center}
\begin{tabular}{@{}l|cr|rrrr@{}}
\toprule
\textbf{Model} & \begin{tabular}[c]{@{}l@{}}\textbf{Sampling} \\   \textbf{Method}\end{tabular} 
 &
  \multicolumn{1}{l|}{\textbf{Steps}} &
  \multicolumn{1}{l}{\textbf{ \begin{tabular}[c]{@{}l@{}}\textbf{Speed} \\   \textbf{(sec$^{-1}$)}\end{tabular} }} &
  \multicolumn{1}{l}{\textbf{Mol Sta \%}} &
  \multicolumn{1}{l}{\textbf{Valid \%}} &
  \multicolumn{1}{l}{\textbf{\begin{tabular}[c]{@{}l@{}}\textbf{Valid \& } \\   \textbf{Unique \%}\end{tabular}}} \\ \midrule
% EDM                        & DDPM                  & 1000 & 3.70   & 82.0         & 91.9         & 90.7         \\ \midrule
\multirow{2}{*}{GeoLDM}    & \multirow{2}{*}{DDPM} & 1000 & 3.70   & 89.4         & 93.8         & 92.7         \\
                           &                       & 100  & 33.30  & 55.8         & 70.6         & 79.7         \\ \midrule
\multirow{3}{*}{GeoLDM}    & \multirow{3}{*}{DDIM} & 1000 & 3.59   & 76.3         & 87           & 86.1         \\
                           &                       & 125  & 28.30  & 74.6         & 85.3         & 84.1         \\
                           &                       & 16   & 196.69 & 31.4         & 53.0         & 52.7         \\ \midrule
\multirow{2}{*}{GeoLDM-PD} & \multirow{2}{*}{DDPM} & 125  & 28.28  & 88.4         & 93.3         & 91.6         \\
                           &                       & 16   & 196.51 & 51.0         & 73.2         & 72.3         \\ \midrule
\multirow{2}{*}{GeoLDM-PD} & \multirow{2}{*}{DDIM} & 125  & 28.28  & 81.6         & 91.7         & 83.6         \\
                           &                       & 16   & 196.51 & 50.4         & 73.4         & 72.6 \\ \bottomrule 
\end{tabular}
\caption{\textbf{Comparison of base model quality and timing based on number of diffusion steps.} Speed is reported as the average number of molecules generated per second. Equivariant latent progressive distillation enables $\sim7.5\times$ speed-up in sample time with minimal loss in sample quality.}
\label{tab:comparison}
\end{center}
\end{table*}

\begin{figure*}[ht]
    \centering
    \includegraphics[width=10cm]{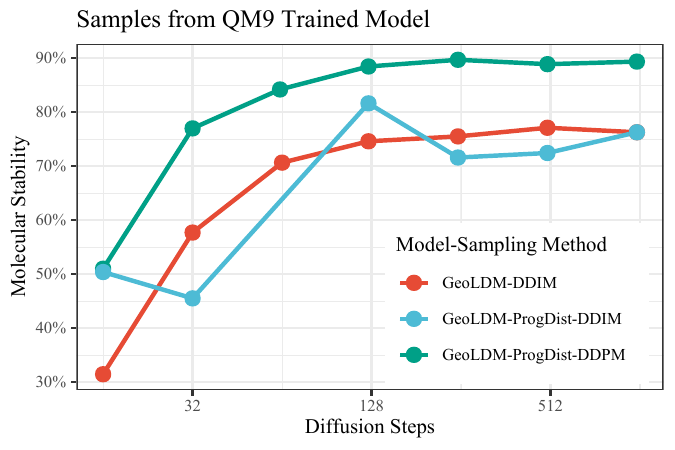}
\caption{\textbf{Molecular stability of generated samples for various diffusion steps.} GeoLDM-DDIM is the original model sampled for a varying number of DDIM steps. GeoLDM-ProgDist-DDIM and GeoLDM-ProgDist-DDPM are sequences of progressively distilled models, using DDIM and DDPM samplers at train time. Note: $x$ axis is logarithmic; distillation progresses from right to left.}
\label{fig:comparison}
\end{figure*}

%%%%%%%%%%%%%%%%%%%%%%%%%%%%%%%%%%%%%%%%%%%%%%%%

\section{Analysis}

\subsection{Accelerated vs Baseline Models}

Compared to the baseline, all distilled models and reduced-step models achieve significant speed-ups in terms of wall-clock time to generate a sample molecule. There are no major differences in speed between the different methods after accounting for the number of diffusion steps. The GeoLDM model that was trained for 100 diffusion steps improves speed but suffers a large decrease in sample quality, with a 33.6\% drop in molecular stability. The models that were trained with 1000 steps and subsequently accelerated all exhibit significantly better quality at similar speed. 

\subsection{Comparing Acceleration Methods}

The deterministic DDIM sampler performs significantly worse than the DDPM sampler even at an equivalent number of denoising steps. There is a small drop in quality when sampling with 125 steps, and a much sharper drop with fewer steps. Because of this immediate drop in quality, DDIM sampling is likely not a viable option for accelerating GeoLDM models. Progressive distillation with deterministic sampling, as proposed in the initial formulation, performs similarly to direct DDIM sampling on the teacher model except at very low numbers of steps.

Despite the fact that the progressive distillation process trains the student model to match DDIM trajectories with the teacher model, we see that performing stochastic DDPM sampling to train the student models offers better sample quality than DDIM sampling until reaching 16 steps (6 halvings). This is consistent with the results in \citet{salimans2022progressive}, who show that for larger numbers of steps, stochastic sampling works better than DDIM sampling. Specifically, equivariant latent progressive distillation (ELPD) with DDPM sampling leads to molecular stability scores higher than regular DDIM, and than ELPD with DDIM, by and average of \textbf{+24.0\%} and \textbf{+26.4\%}, respectively. 

As a conclusion, we find that \textbf{progressive distillation to 125 steps (3 halvings) with stochastic sampling (DDPM) offers the best overall trade-off between speed gains and quality loss, with a roughly 7.5$\times$  speed-up and only a 1 point drop in molecular stability}. Further speed gains can be achieved by performing further distillations, at a trade-off with decreased molecular stability. 

\subsection{Known Limitations}
Compute constraints and the need to sequentially train models for each step of distillation lead us to report results for a single sequence successive distillation training run for each setup, and we report evaluation results for a single distilled model at each step (hence, no error bars). Of note: we encountered numerical instabilities when evaluating models at further stages of training, leading to unstable generations such as with the 63-step DDIM GeoLDM-ProgDist model (see Table \ref{tab:full}). 

Lastly, stability as measured by the valence of atoms is an imperfect measurement of generation quality, as it doesn't take into account conformation energy, likelihood, steric effects, or any other consideration than valence. As a result, some generations show implausible molecular conformations, especially at low numbers of sampling steps (e.g. images (e) and (f) sampled from the 16-step DDPM model after 6 halvings). Further work could incorporate conformation energy estimations to realistically sample the Boltzmann distribution with higher-likelihood generations.

%%%%%%%%%%%%%%%%%%%%%%%%%%%%%%%%%%%%%%%%%%%%%%%%

\section{Conclusion}

In conclusion, our work applies recent advances in fast diffusion model sampling to accelerate 3-dimensional molecular conformations generation by building on the recently introduced GeoLDM  equivariant geometric diffusion model \citep{xu2023geometric}. 

Our primary contribution lies in the introduction and evaluation of Equivariant Latent Progressive Distillation, a diffusion model acceleration technique in latent space which enforces equivariance by performing progressive distillation of the equivariant latent diffusion process of an original GeoLDM pre-trained teacher model via successive halving of diffusion steps. We implement and evaluate speed and quality of latent diffusion models trained through ELPD with both stochastic (DDPM) and deterministic (DDIM) diffusion steps sampling at train and inference time.

Our findings indicate that stochastic sampling (DDPM) of equivariant latent diffusion models trained through progressive distillation maintain molecular structure quality significantly better than other fast sampling methods at comparable sampling speed-ups. \textbf{These results suggest that the Equivarient Latent Progressive Distillation method has strong potential for high-throughput \emph{in silico} molecular conformations screening.}

Overall, our research provides a deeper understanding of the trade-offs between sampling speed gains and quality losses in fast sampling of generative models for molecular conformations. The insights and methods presented in this paper pave the way for faster molecular structures generation of large bio-molecules, which are essential to computational biochemistry, drug discovery, and life sciences.

%%%%%%%%%%%%%%%%%%%%%%%%%%%%%%%%%%%%%%%%%%%%%%%%

\subsubsection*{Acknowledgments}
The authors wish to thank Stefano Ermon, Ron Dror, and Minkai Xu at Stanford University for their help, and for the GeoLDM model and implementation which this work extends.

\subsection*{Code and Data}
Code and data for our experiments are available on the following GitHub repository: 

\url{https://github.com/rlacombe/AccGeoLDM}.

\subsection*{Future Work}

In future work, these results could be extended to larger molecules by experimenting with similar models and training setups on the \emph{GEOM-DRUG} (Geometric Ensemble Of Molecules) dataset introduced by \citet{Axelrod2022GEOM}, which features larger organic compounds.

Also, the GeoLDM implementation uses a `light' VAE where only the decoder is trained in practice. Further experiments could explore whether richer variational representations of the input molecules in physical space could lead to higher structural stability for molecules sampled from latent diffusion. 

Additionally, a GeoLDM model could be trained with a continuous-time denoising diffusion model \citep{karras2022elucidating} in latent space, or other continuous-time accelerated sampling methods such as consistency models \citep{song2023consistency}.

Lastly, stability of molecules as measured by the valence of generated atoms is an imperfect measurement of the quality of conformations.  Further work could incorporate conformation energy estimations to realistically sample the Boltzmann distribution with higher-likelihood generations.

\bibliography{iclr2024_conference}
\bibliographystyle{iclr2024_conference}

\appendix

%%%%%%%%%%%%%%%%%%%%%%%%%%%%%%%%%%%%%%%%%%%%%%%%

\section{Appendix: Detailed Results} 
We report the detailed results of our experiences with fast sampling methods in Table \ref{tab:full}.

\begin{figure}[ht]
    \centering
    \begin{subfigure}[b]{0.46\textwidth}
        \centering
        \includegraphics[width=\textwidth]{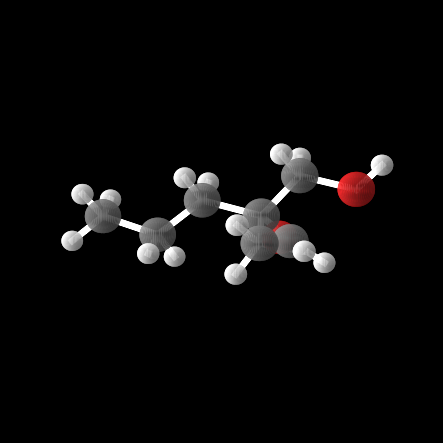 }
        \caption{ }
    \end{subfigure}
    \hspace{0.5cm}
    \begin{subfigure}[b]{0.46\textwidth}
        \centering
        \includegraphics[width=\textwidth]{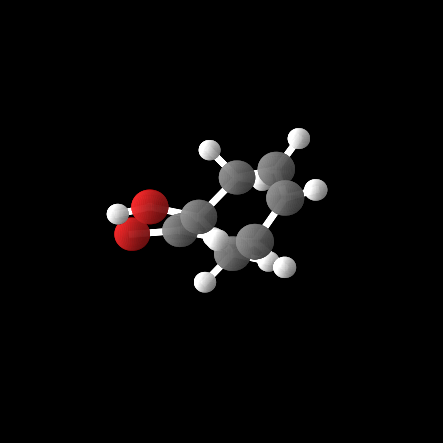 }
        \caption{}
    \end{subfigure}

    \begin{subfigure}[b]{0.46\textwidth}
        \centering
        \includegraphics[width=\textwidth]{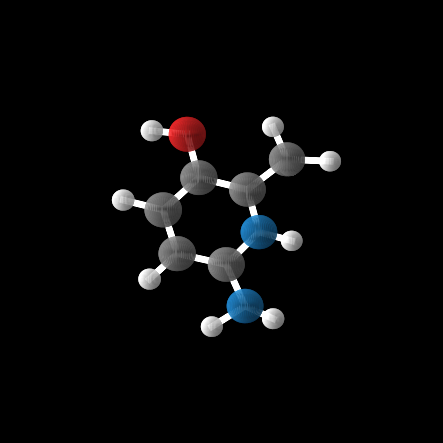 }
        \caption{}
    \end{subfigure}
    \hspace{0.5cm}
    \begin{subfigure}[b]{0.46\textwidth}
        \centering
        \includegraphics[width=\textwidth]{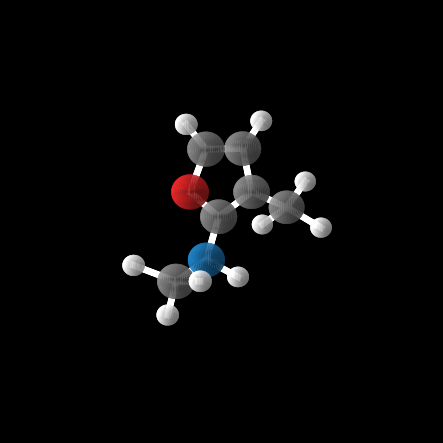 }
        \caption{}
    \end{subfigure}

    \begin{subfigure}[b]{0.46\textwidth}
        \centering
        \includegraphics[width=\textwidth]{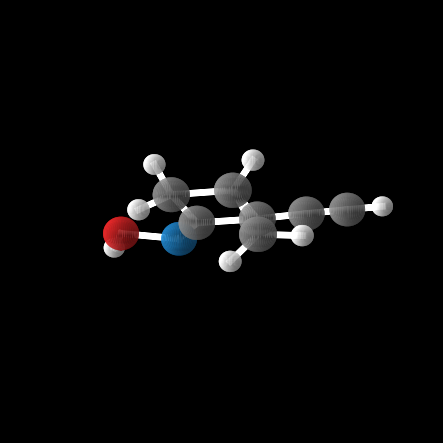 }
        \caption{}
    \end{subfigure}
    \hspace{0.5cm}
    \begin{subfigure}[b]{0.46\textwidth}
        \centering
        \includegraphics[width=\textwidth]{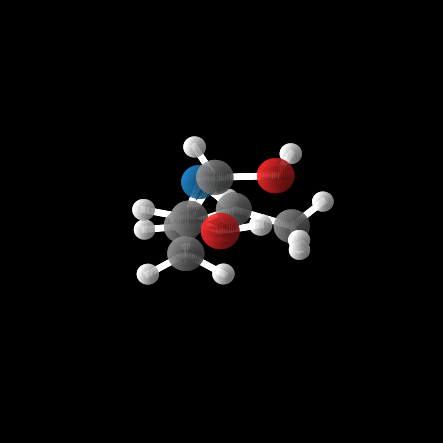 }
        \caption{}
    \end{subfigure}
    \caption{\textbf{Examples of conformation generations}, sampled from: (a, b) original 1000-step GeoLDM model; (c, d) 125-step model distilled with DDPM sampler after 3 halvings; (e, f) 16-step model distilled with DDPM sampler after 6 halvings.}
    \label{fig:visualizations}
\end{figure}

We also report a selection of example visualizations of conformations generated by various models in Figure \ref{fig:visualizations} (next page).

\begin{table}[ht]
\centering
\resizebox{\textwidth}{!}{%
\begin{tabular}{l|lr|rrrrr}
\hline
\textbf{Model} &
  \textbf{ \begin{tabular}[c]{@{}l@{}}\textbf{Sample} \\   \textbf{Method}\end{tabular}} &
  \multicolumn{1}{l}{\textbf{Steps}} &
  \multicolumn{1}{l}{\textbf{ \begin{tabular}[c]{@{}l@{}}\textbf{Speed} \\   \textbf{(sec$^{-1}$)}\end{tabular} }} &
  \multicolumn{1}{l}{\textbf{Mol Sta (\%)}} &
  \multicolumn{1}{l}{\textbf{Atom Sta (\%)}} &
  \multicolumn{1}{l}{\textbf{Valid (\%)}} &
  \multicolumn{1}{l}{ \begin{tabular}[c]{@{}l@{}}\textbf{Valid \&} \\   \textbf{Unique (\%)}\end{tabular} }\\ \hline
EDM                              & DDPM                   & 1000 & 3.70                 & 82.00 & 98.70 & 91.90 & 90.70 \\ \hline
\multirow{2}{*}{GeoLDM}          & \multirow{2}{*}{DDPM}  & 1000 & 3.70                 & 89.40 & 98.90 & 93.80 & 92.70 \\
                                 &                        & 100  & 33.33                & 55.80 & 95.40 & 70.60 & 79.70 \\ \midrule
\multirow{11}{*}{GeoLDM}         & \multirow{11}{*}{DDIM} & 1000 & 3.59                 & 76.30 & 97.00 & 87.00 & 86.06 \\
                                 &                        & 500  & 7.14                 & 77.14 & 97.13 & 87.10 & 85.92 \\
                                 &                        & 250  & 14.16                & 75.54 & 96.90 & 85.91 & 84.96 \\
                                 &                        & 125  & 28.40                & 74.63 & 96.66 & 85.37 & 84.16 \\
                                 &                        & 63   & 54.3                 & 70.67 & 95.95 & 82.47 & 81.48 \\
                                 &                        & 32   & 105.06               & 57.71 & 93.43 & 74.91 & 74.09 \\
                                 &                        & 16   & 196.69               & 31.46 & 85.83 & 53.05 & 52.75 \\
                                 &                        & 8    & 349.04               & 3.67  & 66.33 & 20.36 & 19.96 \\
                                 &                        & 4    & 569.16               & 0.00  & 31.87 & 13.31 & 2.96  \\
                                 &                        & 2    & 833.26               & 0.00  & 1.29  & 75.38 & 0.14  \\
                                 &                        & 1    & 1097.26              & 0.00  & 0.14  & 97.93 & 0.06  \\ \midrule
\multirow{7}{*}{GeoLDM-ProgDist} & \multirow{7}{*}{DDPM}  & 1000 & 3.70                 & 89.40 & 98.90 & 93.80 & 92.70 \\
                                 &                        & 500  & 7.17                 & 88.93 & 98.85 & 93.23 & 91.56 \\
                                 &                        & 250  & 14.31                & 89.74 & 98.91 & 93.53 & 92.01 \\
                                 &                        & 125  & 28.28                & 88.50 & 98.78 & 93.38 & 91.70 \\
                                 &                        & 63   & 55.18                & 84.24 & 98.20 & 90.89 & 89.15 \\
                                 &                        & 32   & 104.77               & 77.02 & 97.24 & 87.23 & 86.19 \\
                                 &                        & 16   & 196.51               & 51.02 & 92.08 & 73.22 & 72.39 \\ \midrule
\multirow{7}{*}{GeoLDM-ProgDist} & \multirow{7}{*}{DDIM}  & 1000 & 3.59                 & 76.30 & 97.00 & 87.00 & 86.06 \\
                                 &                        & 500  & 7.17                 & 72.46 & 96.27 & 85.74 & 85.54 \\
                                 &                        & 250  & 14.31                & 71.62 & 96.45 & 82.22 & 81.23 \\
                                 &                        & 125  & 28.28                & 81.66 & 97.62 & 91.71 & 83.67 \\
                                 &                        & 63   & 55.18                & 0.00  & 0.00  & 0.00  & 0.00  \\
                                 &                        & 32   & 104.77               & 45.52 & 91.51 & 58.52 & 57.60 \\
                                 &                        & 16   & 196.51               & 50.40 & 92.07 & 73.44 & 72.65 \\ \bottomrule
\end{tabular}
}
\caption{Full comparison of model quality and timing based on different numbers of diffusion steps. All metrics are reported on the QM9 dataset.}
\label{tab:full}
\end{table}

\newpage

\clearpage

%%%%%%%%%%%%%%%%%%%%%%%%%%%%%%%%%%%%%%%%%%%%%%%%

\section{Appendix: Equivariant Latent Progressive Distillation Algorithm}

Here we present the detailed Equivariant Latent Progressive Distillation algorithm introduced in the Methods section and represented in Figure \ref{fig:progressive-distillation}.

\begin{algorithm}[ht]
\caption{Equivariant Latent Progressive Distillation:}\label{alg:progdist}
\begin{algorithmic}[1]
\Require Trained teacher model $\hat{z}_\eta(z_t)$
\Require Data set $D$
\Require Loss weight function $w()$
\Require Student sampling steps $N$
\Require Noise schedule $\sigma_t, \alpha_t \, \forall t \in [1,N]$
\For{$K$ iterations}
    \State $\theta \leftarrow \eta$ \Comment{Init student from teacher}
    \While{not converged}
        \State $x, h \sim D$
        \State \textbf{Without gradient computation:}
        \Comment{VAE and teacher are fixed} 
        \State \hspace{2em} $t = i/N, i \sim \text{Cat}[1, 2, \ldots, N]$
    \State \hspace{2em} $z_{xh} = \mathcal{E}_\phi(x, h)$ 
    \Comment{Encode $x, h$ in latent space}
    \State \hspace{2em} $\epsilon \sim \mathcal{N}(0, I)$
        \State \hspace{2em} $z_t = \alpha_t z_{xh} + \sigma_t \epsilon$
        \State \hspace{2em} $t' = t - 0.5/N, t'' = t - 1/N$ \Comment{Take 2 sampling steps with Teacher}
        \State \hspace{2em} $z_{t'} = \alpha_{t'} \hat{z}_\eta(z_t) + \sigma_{t'} \frac{\sigma_{t'}}{\sigma_t}(z_t - \alpha_t \hat{z}_\eta(z_t))$
        \State \hspace{2em} $z_{t''} = \alpha_{t''} \hat{z}_\eta(z_{t'}) + \sigma_{t''} \frac{\sigma_{t''}}{\sigma_{t'}}(z_{t'} - \alpha_{t'} \hat{z}_\eta(z_{t'}))$
        \State \hspace{2em} $\tilde{z} = \frac{z_{t''} - (\sigma_{t''}/\sigma_t)z_t}{\alpha_{t''} - (\sigma_{t''}/\sigma_t)\alpha_t}$ \Comment{Teacher $\tilde{z}$ latent target}
        \State \hspace{2em}\hspace{2em} $\lambda_t = \log[\alpha_t^2 / \sigma_t^2]$
    \State Compute $\hat{z}_\theta(z_t)$ 
    \Comment{Student $\hat{z}_\theta$ latent prediction}
    \State $L_\theta =  w(\lambda_t)\|\tilde{z} - \hat{z}_\theta(z_t)\|^2$
    \State $\theta \leftarrow \theta - \gamma \nabla_\theta L_\theta$
    \Comment{Gradient descent step to update student}
    \EndWhile
    \State $\eta \leftarrow \theta$ \Comment{Student becomes next teacher}
    \State $N \leftarrow \lceil N/2 \rceil$  \Comment{Halve number of sampling steps}
\EndFor
\end{algorithmic}
\end{algorithm}

\end{document}